\documentclass[letterpaper,twocolumn,prl,aps,superscriptaddress,amsmath,amssymb,floatfix]{revtex4-1}
\usepackage{mathptmx}
\usepackage[utf8]{inputenc} 
\usepackage[T1]{fontenc}    
\usepackage{color}
\usepackage{amsmath}
\usepackage{amssymb}
\usepackage{graphicx}
\usepackage{esint}
\usepackage[unicode=true,
 bookmarks=true,bookmarksnumbered=false,bookmarksopen=false,
 breaklinks=false,pdfborder={0 0 1},backref=false,colorlinks=true]
 {hyperref}
\hypersetup{
 linkcolor=magenta,urlcolor=blue,citecolor=blue,pdfstartview={FitH},hyperfootnotes=false}

\makeatletter

\usepackage{textcomp}
\usepackage{epstopdf}

\pdfpageheight\paperheight
\pdfpagewidth\paperwidth

\@ifundefined{textcolor}{}{%
 \definecolor{BLACK}{gray}{0}
 \definecolor{WHITE}{gray}{1}
 \definecolor{RED}{rgb}{1,0,0}
 \definecolor{GREEN}{rgb}{0,1,0}
 \definecolor{BLUE}{rgb}{0,0,1}
 \definecolor{CYAN}{cmyk}{1,0,0,0}
 \definecolor{MAGENTA}{cmyk}{0,1,0,0}
 \definecolor{YELLOW}{cmyk}{0,0,1,0}
}

\usepackage{xcolor}\usepackage{soul}
\newcommand{\ket}[1]{\ensuremath{\left|#1\right\rangle}}

\definecolor{blue}{rgb}{0,0,1}
\definecolor{red}{rgb}{1,0,0}
\definecolor{green}{rgb}{0,1,0}

\usepackage{siunitx}
\usepackage{physics} 
\usepackage{mathrsfs}

\usepackage{bm}
\usepackage{mhchem}

\usepackage{color}
\usepackage{ulem}

\renewcommand{\i}{\mathrm{i}}

\makeatother

\begin{document}
\title{Proposal for Forward Brillouin Inter-Modal Scattering in Non-suspended Lithium Niobate Waveguides at Visible Wavelengths}

\author{Jia-Lin Chen}
\affiliation{Laboratory of Quantum Information, 
University of Science and Technology of China, Hefei 230026, China}
\affiliation{Anhui Province Key Laboratory of Quantum Network, 
University of Science and Technology of China, Hefei 230026, China}

\author{Yuan-Hao Yang}
\affiliation{Laboratory of Quantum Information, 
University of Science and Technology of China, Hefei 230026, China}
\affiliation{Anhui Province Key Laboratory of Quantum Network, 
University of Science and Technology of China, Hefei 230026, China}

\author{Zheng-Xu Zhu}
\affiliation{Laboratory of Quantum Information, 
University of Science and Technology of China, Hefei 230026, China}
\affiliation{Anhui Province Key Laboratory of Quantum Network, 
University of Science and Technology of China, Hefei 230026, China}

\author{Jia-Qi Wang}
\affiliation{Laboratory of Quantum Information, 
University of Science and Technology of China, Hefei 230026, China}
\affiliation{Anhui Province Key Laboratory of Quantum Network, 
University of Science and Technology of China, Hefei 230026, China}

\author{Xin-Biao Xu}
\affiliation{Laboratory of Quantum Information, 
University of Science and Technology of China, Hefei 230026, China}
\affiliation{Anhui Province Key Laboratory of Quantum Network, 
University of Science and Technology of China, Hefei 230026, China}

\author{Ming Li}
\affiliation{Laboratory of Quantum Information, 
University of Science and Technology of China, Hefei 230026, China}
\affiliation{Anhui Province Key Laboratory of Quantum Network, 
University of Science and Technology of China, Hefei 230026, China}
\affiliation{CAS Center for Excellence in Quantum Information and Quantum Physics, University of Science and Technology of China, Hefei 230026, China}
\affiliation{Hefei National Laboratory, University of Science and Technology of China, Hefei 230088, China}

\author{Zheng-Fu Han}
\affiliation{Laboratory of Quantum Information, 
University of Science and Technology of China, Hefei 230026, China}
\affiliation{Anhui Province Key Laboratory of Quantum Network, 
University of Science and Technology of China, Hefei 230026, China}
\affiliation{CAS Center for Excellence in Quantum Information and Quantum Physics, University of Science and Technology of China, Hefei 230026, China}
\affiliation{Hefei National Laboratory, University of Science and Technology of China, Hefei 230088, China}

\author{Guang-Can Guo}
\affiliation{Laboratory of Quantum Information, 
University of Science and Technology of China, Hefei 230026, China}
\affiliation{Anhui Province Key Laboratory of Quantum Network, 
University of Science and Technology of China, Hefei 230026, China}
\affiliation{CAS Center for Excellence in Quantum Information and Quantum Physics, University of Science and Technology of China, Hefei 230026, China}
\affiliation{Hefei National Laboratory, University of Science and Technology of China, Hefei 230088, China}

\author{Wei Chen}
\email{weich@ustc.edu.cn}
\affiliation{Laboratory of Quantum Information, 
University of Science and Technology of China, Hefei 230026, China}
\affiliation{Anhui Province Key Laboratory of Quantum Network, 
University of Science and Technology of China, Hefei 230026, China}
\affiliation{CAS Center for Excellence in Quantum Information and Quantum Physics, University of Science and Technology of China, Hefei 230026, China}
\affiliation{Hefei National Laboratory, University of Science and Technology of China, Hefei 230088, China}

\author{Chang-Ling Zou}
\email{clzou321@ustc.edu.cn}
\affiliation{Laboratory of Quantum Information, 
University of Science and Technology of China, Hefei 230026, China}
\affiliation{Anhui Province Key Laboratory of Quantum Network, 
University of Science and Technology of China, Hefei 230026, China}
\affiliation{CAS Center for Excellence in Quantum Information and Quantum Physics, University of Science and Technology of China, Hefei 230026, China}
\affiliation{Hefei National Laboratory, University of Science and Technology of China, Hefei 230088, China}

\date{\today}

\begin{abstract}
Thin-film lithium niobate on sapphire provides an excellent platform for simultaneously confining acoustic and optical modes without suspended structures, enabling efficient acousto-optic modulation through strong piezoelectric coupling. Here, we identify the challenges in realizing the forward Brillouin interaction at visible wavelengths, and overcome the limitation by introducing a quasi-phase-matching scheme through periodic waveguide width modulation. We predict a complete inter-modal optical conversion over \SI{1.1}{mm} using only \SI{1}{mW} acoustic power. Our study paves the way for high-performance visible-wavelength acousto-optic devices on integrated platforms.
\end{abstract}
\maketitle

\section{Introduction}

Acousto-optic (AO) devices have emerged as critical components for on-chip photonic systems, offering unique advantages over competing modulation technologies. The AO devices employes the photo-elastic interaction which manifest one of the most strongest nonlinear optics interaction in dielectrics, thus promises more compact footprint or lower pump powers compared to other optics modulators relying on the electro-optic~\cite{wang_Integrated_2018,xue_Breaking_2022,sinatkas_Electrooptic_2021} or thermo-optic~\cite{watts_Adiabatic_2013,lu_Michelson_2015,yu_Compact_2022} effects. 
Recently, the integrated AO devices with compelling practical benefits, such as device footprints below \SI{0.1}{mm^2}, power consumption in the milliwatt range, and modulation speeds reaching hundreds of megahertz, have been demonstrated experimentally~\cite{zhang_Highly_2025,xu_Highfrequency_2022,liu_Electromechanical_2019}. The demonstrated integrated AO devices can be divided into two primary configurations: side-coupled designs where surface acoustic waves interact with guided optical modes from the lateral direction~\cite{kittlaus_Electrically_2021,sarabalis_Acoustooptic_2020,zhang_Onchip_2025}, and co-propagating architectures where both acoustic and optical waves are confined within the same waveguide structure~\cite{sarabalis_Acoustooptic_2021,zhang_Highly_2025,Xu2025, Yang2025}. The latter approach has recently achieved remarkable success through backward Brillouin scattering demonstrations in thin-film lithium niobate (LN), where the material's exceptional piezoelectric properties enable microwave-to-acoustic transduction efficiencies exceeding 10\% at 9~GHz frequencies~\cite{Xu2025,Yang2025}, combined with simultaneous confinement of photons and phonons in the same subwavelength waveguide.

\begin{figure}[ht!]
    \centering
    \includegraphics[width=.75\columnwidth]{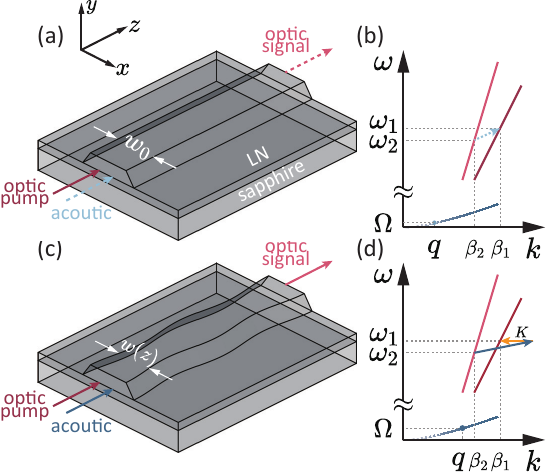}
    \caption{(a) Waveguide structure schematic and (b) phase-matching condition illustration of forward inter-modal Brillouin scattering. (c) Waveguide structure schematic and (d) phase-matching condition illustration of proposed QPM forward Brillouin scattering scheme.}
    \label{fig-1}
\end{figure}

However, backward Brillouin scattering faces fundamental limitations when extended to visible wavelengths, a spectral regime of critical importance for quantum technologies~\cite{Lu2024}, particularly the manipulation of neutral atoms by the optical transition at visible wavelengths, such as the D2 transition of Rubidium atoms at \SI{780}{nm}~\cite{Browaeys2020}. At these visible wavelengths, momentum conservation demands acoustic frequencies exceeding \SI{15}{GHz} with corresponding acoustic wavelengths below \SI{200}{nm}~\cite{yang_Proposal_2024}. Fabricating interdigital transducers (IDTs) for such high frequencies approaches the limits of current fabrication technology, and acoustic losses become prohibitively high at these frequencies.
Forward Brillouin scattering offers an alternative pathway, requiring acoustic frequencies in the more manageable range typically below \SI{1}{GHz}~\cite{sarabalis_Acoustooptic_2021,zhang_Integratedwaveguidebased_2024,zhang_Highly_2025}. This approach maintains the advantage of simultaneous photon-phonon confinement within a single waveguide while operating at frequencies readily accessible through conventional IDT fabrication, providing a potentially viable route to efficient AO interaction and modulation at visible wavelengths.

In this work, we identify and resolve a fundamental challenge in designing forward Brillouin scattering devices at visible wavelengths. The core dilemma lies in a trade-off between acoustic confinement and phase-matching: achieving tight acoustic confinement requires short acoustic wavelengths (high frequencies), yet phase-matching between optical modes demands long acoustic wavelengths (low frequencies) due to the small propagation constant differences between guided optical modes. This contradiction has prevented the realization of efficient forward Brillouin devices in non-suspended visible-wavelength waveguides. We introduce quasi-phase-matching (QPM) through periodic waveguide width modulation to break this fundamental constraint. By providing additional momentum through the engineered grating structure, QPM enables the use of well-confined, higher-frequency acoustic modes while maintaining the phase-matching condition. Our optimized design achieves a large coupling coefficient comparable to state-of-the-art infrared devices, with complete modal conversion predicted over just 1.1 mm using 1 mW of acoustic power. This approach not only solves the immediate challenge but also opens new design freedoms for engineering AO interactions across the visible spectrum.

\begin{figure}[ht!]
    \centering
    \includegraphics[width=.99\columnwidth]{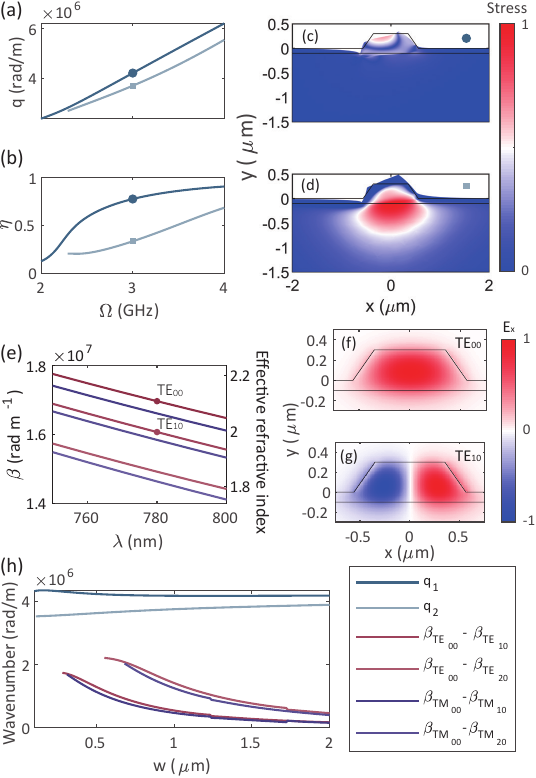}
    \caption{(a) Dispersion curve and (b) confinement factor of acoustic modes. (c) and (d) are profiles of two acoustic modes annoted with circle and square in (a) and (b). (e) Dispersion curves of optical modes: TE mode (red) and TM mode (blue). (f) and (g) are TE$_{00}$ and TE$_{01}$ mode optic profiles. (h) Dependence of acoustic mode wavenumbers and optical modes propagation constant differences on the waveguide width.}
    \label{fig-2}
\end{figure}

Figure~\ref{fig-1}(a) depicts forward inter-mode Brillouin scattering in a LN ridge waveguide on sapphire substrate. This platform configuration provides simultaneous confinement of acoustic and optical modes, exploiting the lower acoustic and optical velocities in the LN relative to the substrate, combined with strong piezoelectricity that enable efficient RF to acoustic wave transduction. It is expected to realize the strong interaction between co-propagating optical and acoustic waves~\cite{yang_Stimulated_2023}. As illustrated in Fig.~\ref{fig-1}(b), by input an optical pump ($\omega_1$, $\beta_1$) and an acoustic wave ($\Omega$, $q$), a frequency-shifted output optical signal ($\omega_1$, $\beta_1$) can be generated, with their frequencies satisfying energy conservation condition $\omega_2=\omega_1+\Omega$ and their propagation wavenumber satisfying the phase matching condition $\beta_2=\beta_1+q$. However, one significant challenge arises when implement this concept in TFLN waveguides: co-directional optical modes has small wavenumber differences on the order of typically at order of \SI{E6}{rad/m}, which corresponds to acoustic wavelengths in the micrometer range that could not be effectively confined within the waveguide. 

To quantify this challenge, we examine a representative waveguide structure fabricated from $400$-\unit{nm}-thick {x-cut} LN film on sapphire substrate with a top width of \SI{0.7}{\micro m}, an etch depth of \SI{300}{nm}, and a sidewall angle of \SI{55}{\degree}. The waveguide structure {is along z axis of the LN and} supports {at most} two acoustic modes, as shown in Fig.~\ref{fig-2}(a). The confinement factors ($\eta$), defined as the fraction of acoustic power flow confined within the LN ridge region, increases at higher acoustic frequency (Fig.~\ref{fig-2}(b)), indicating stronger mode confinement for shorter wavelengths. For acoustic mode at low frequency such as \SI{2.1}{GHz}, the wavenumber reaches \SI{2.5E6}{rad/m} with weak confinement $\eta$ of only $18.6\%$, which would significantly reduce AO coupling. On the other hand, acoustic modes at higher frequency such as \SI{3}{GHz} (Fig.~\ref{fig-2}(c) and (d)), show better confinement and greater wavenumbers. However, Fig.~\ref{fig-2}(e) reveals that the propagation constant difference of any two optic modes in this waveguide is no more than \SI{2.34E6}{rad/m}. In particular, the propagation constant of TE$_{00}$ (Fig.~\ref{fig-2}(f)) and TE$_{10}$ (Fig.~\ref{fig-2}(g)) modes is only \SI{0.90E6}{rad/m}. Consequently, the phase matching condition cannot be directly satisfied for these well-confined acoustic modes (Figs.~\ref{fig-2}(c)-(d)). {This fundamental limitation remains unresolved across different waveguide dimensions. As shown in Fig.~\ref{fig-2}(h), while the propagation constant difference of the optical modes increases with narrowing waveguides, the optical modes cease to be supported below certain dimensions, with the maximum optical propagation constant difference reaching only \SI{2.22E6}{rad/m}. For comparison, the wavenumbers of acoustic modes exhibit less sensitivity to waveguide width, maintaining values above \SI{3.5E6}{rad/m}. Consequently, no waveguide width exists that simultaneously satisfies phase matching conditions for both \SI{3}{GHz} acoustic modes and \SI{780}{nm} optical modes. Further simulations (shown in supplemental document) indicate that this issue persists across varying LN film thicknesses and etch depths, fundamentally precluding simultaneous achievement of strong acoustic confinement and phase-matching conditions.}

To overcome such dilemma, we propose to modulate the width of the waveguide to realize the QPM, as depicted in Fig.~\ref{fig-1}(c). For a modulation period of $\Lambda$,  the optical permittivity, elasto-optic coefficient and electro-optic coefficient distribution are modulated along $z$, providing an extra momentum of $K=\frac{2\pi}{\Lambda}$ to the modes, as shown in Fig.~\ref{fig-1}(d). In the following simulation, we consider the coupling between TE$_{00}$ (Fig.~\ref{fig-2}(f)) and TE$_{10}$ (Fig.~\ref{fig-2}(g)) modes with \SI{3}{GHz} acoustic mode (Fig.~\ref{fig-2}(c)), and we modulate the waveguide width in the sinusoidal form: $w(z)=w_0+\Delta w\sin(2\pi z/\Lambda)$ with a modulation amplitude $\Delta w=$ \SI{0.1}{\micro m}. The AO coupling coefficient is approximately in the form of $g=g_0+\frac{1}{2}g_{\text{eff}}\sin(2\pi z/\Lambda)$, and the effective coupling coefficient $g_{\text{eff}}$ is estimated as the difference of coupling coefficients in the narrowest and widest cross-section $g_{\text{eff}}=g_{\text{narrow}}-g_{\text{wide}}$. 

\begin{figure}[t!]
    \centering
    \includegraphics[width=.95\columnwidth]{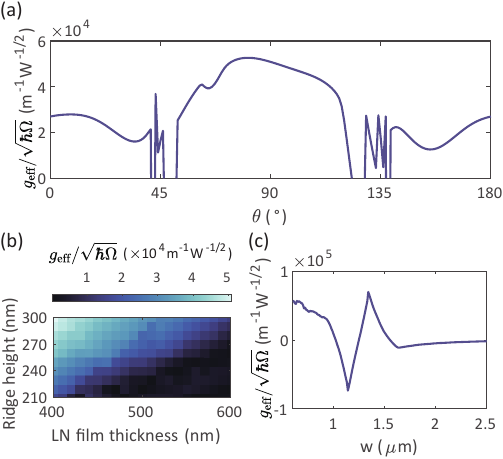}
    \caption{(a) Coupling coefficient under different propagation direction. $\theta$ is the angle between propagation direction and LN $+y$ axis and $\theta=$ \SI{90}{\degree} corresponds to propagating along LN $+z$ axis. (b) Coupling coefficient under different LN film thickness and etch depth. (c) Coupling coefficient under different average waveguide width.}
    \label{fig-3}
\end{figure}

As shown in Fig.~\ref{fig-3}(a), $g_{\text{eff}}$ varies with the propagation direction of the waveguide ($\theta$) due to the anisotropy of LN, and we choose $\theta=$ \SI{90}{\degree}, i.e. propagation along LN $z$-axis to achieve a reasonably strong coupling coefficient. Notably, the coupling coefficient exhibits significant fluctuations near $\theta=$ \SI{45}{\degree} and \SI{135}{\degree}, which arises from the hybridization of TE and TM modes at these angles. In some angular range, $g_{\text{eff}}$ becomes negative, indicating that the coupling coefficient in the wide waveguide exceeds that in the narrow waveguide. The LN film thickness and etch depth also significantly influence the coupling coefficient. While thicker LN films and deeper etching improve mode confinement, they also reduce the mode power density. As shown in Fig.~\ref{fig-3}(b), {$g_{\text{eff}}$} generally increases with thinner LN films and deeper etching within the simulated parameter range. {To maximize the coupling coefficient, we selected an LN film thickness of \SI{400}{nm} and an etch depth of \SI{300}{nm}}. Figure~\ref{fig-3}(c) illustrates the variation of $g_{\text{eff}}$ with respect to the average waveguide width $w_0$. For narrow $w_0$ situation, $g_{\text{eff}}$ is generally higher in narrower waveguide due to higher mode power density, and we choose an average waveguide width $w_0=$ \SI{0.7}{\micro m}. A significant fluctuation occurs near $w_0=$ \SI{1.2}{\micro m}, which is also attributed to the hybridization of the {TE$_{10}$} and {TM$_{00}$} modes.

\begin{figure}[t!]
    \centering
    \includegraphics[width=.95\columnwidth]{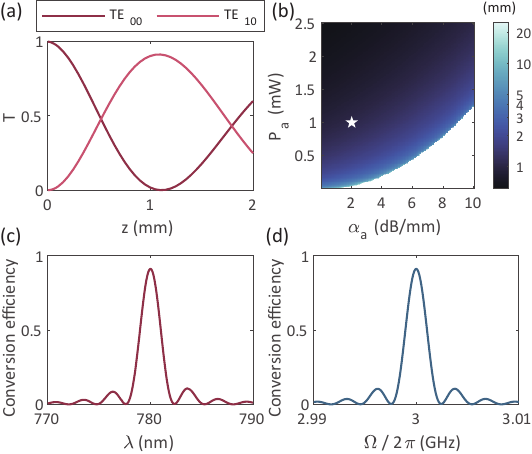}
    \caption{(a) Pump optic mode and signal optic mode transmission about interaction length. (b) Complete inter-modal conversion length under different acoustic mode power and acoustic propagation loss. The white area represent parameter space where complete conversion can not be realized regardless of the interaction length. The star represent the conversion length under parameter combination in (a). Optic mode conversion efficiency as a function of optic wavelength (c) and acoustic frequency (d).}
    \label{fig-4}
\end{figure}

\begin{table*}[htbp]
\centering
\caption{\bf Figures of merit for state-of-art forward brillouin scattering on photonic chips.}
\begin{tabular}{cccccccc}
\hline
Ref. & Year & Material & $\lambda$ (\unit{nm}) & $\Omega/2\pi$ (\unit{GHz}) & $g/\sqrt{\hbar\Omega}$ (\unit{m^{-1}W^{-1/2}}) & Suspended & Acoustic/Optic Confinement\\
\hline
\cite{sarabalis_Acoustooptic_2021} & 2021 & LN & \num{1550} & \num{0.44} & \num{3.77E5} & Y & Y/Y \\
\cite{zhang_Integratedwaveguidebased_2024} & 2024 & GaN & \num{1551.7} & \num{0.998} & \num{2.55E5} & N & Y/Y \\
\cite{zhang_Highly_2025} & 2025 & GaN & \num{1550} & \num{0.85} & \num{1.82E5} & N & Y/Y \\
This work & & LN & \num{780} & $3$ & \num{5.05E4} & N & Y/Y \\
\hline
\end{tabular}
\label{tab:FOM}
\end{table*}

Under these optimized geometry, the predicted $g_{\text{eff}}/\sqrt{\hbar\Omega}=$ \SI{5.05E4}{m^{-1}W^{-1/2}}. To evaluate the practical performance of this design, we consider realistic parameters drawn from recent experimental demonstrations~\cite{Xu2025,Yang2025}. For an IDT excitation efficiency of \SI{10}{\%} and an input RF power of {\SI{10}{mW}}, the acoustic power at the input port is $P_{\text{aco}}(z=0)=$ \SI{1}{mW}. For optical and acoustic modes has propagation losses of {$\alpha_{\text{opt}}=$} \SI{0.37}{dB/mm} and {$\alpha_{\text{aco}}=$} \SI{2.0}{dB/mm}, respectively, corresponding resonance quality factors of {\num{2E5} and \num{9E3}}, complete pump-to-signal mode conversion occurs over a compact waveguide length~\cite{zhang_Highly_2025} of 
\begin{equation*}
    L=-\frac{2}{\alpha_{\text{aco}}}\log\left(1-\frac{\pi\alpha_{\text{aco}}\sqrt{\hbar\Omega}}{4g_{\text{eff}}\sqrt{P_{\text{aco}}(z=0)}}\right)=1.1\,\mathrm{mm},
\end{equation*}
as demonstrated in Fig.~\ref{fig-4}(a).

This milimeter-scale interaction length are beneficial for practical implementations. First, the optical loss is minimized with an insertion loss of only \SI{-0.39}{dB} (\SI{91.4}{\%} power conversion efficiency). Furthermore, the short interaction length relaxes the phase-matching bandwidth constraints, yielding broad 3-dB bandwidths of \SI{2.0}{nm} in the optical domain and \SI{2.1}{MHz} in the acoustic domain. This wide optical bandwidth is critical for addressing Rb D1/D2 line transitions at fixed wavelengths. The acoustic bandwidth arises from the \unit{\micro s}-scale delay for acoustic wave filling the whole waveguide length after IDT excitation. 

The undulating waveguide structure may introduce additional acoustic propagation loss, as the periodic modulation acts as an acoustic grating, scattering the waveguide acoustic mode into bulk acoustic waves in the substrate. Excessive acoustic loss can lead to rapid decay of the acoustic mode, and thus the AO interaction strength is reduced at a given input RF power, resulting in a longer or even infinite waveguide length for unit conversion, as illustrated in Fig.~\ref{fig-4}(b).  For instance, if the acoustic loss reaches \SI{8}{dB/mm}, the required conversion length extends to \SI{2.5}{mm} while the bandwidths are reduced to \SI{1.2}{nm} (optical) and \SI{1.2}{MHz} (acoustic). Additionally, the optical insertion loss will be degraded to \SI{-2.5}{dB}. If the acoustic loss exceeds {the threshold of $\frac{4g_{\text{eff}}\sqrt{P_{\text{aco}}(z=0)}}{\pi\sqrt{\hbar\Omega}}=$} \SI{8.9}{dB/mm}, the unit conversion length becomes infinite, making complete conversion unachievable regardless of the interaction length. {To maintain a short interaction length while compensating for high acoustic propagation losses, increasing the on-chip acoustic power provides an effective solution that simultaneously preserves bandwidth and insertion loss characteristics.} For example, even when operating with an acoustic loss of \SI{8.9}{dB/mm} that at the threshold, an elevated acoustic power of \SI{2.2}{mW} enables maintenance of the desired \SI{1.1}{mm} conversion length while avoiding performance degradation in both spectral bandwidth and conversion efficiency.

Our design achieves an AO coupling strength comparable to devices operating in the telecommunication band, as shown in Table~\ref{tab:FOM}. Compared with AO modulation in GaN waveguide~\cite{zhang_Integratedwaveguidebased_2024,zhang_Highly_2025}, our work has comparable coupling coefficient. Additionally, when considering the strong piezoelectric effect of LN, acoustic wave excitation in LN is more efficient~\cite{ogi_Acoustic_2002}, resulting in lower power consumption. In comparison to other works utilizing suspended LN waveguides~\cite{sarabalis_Acoustooptic_2021}, our design exhibits a lower coupling coefficient, as only a portion of the waveguide cross-section contributes to the AO coupling. However, the non-suspended waveguide structure offers greater mechanical robustness, enabling higher acoustic mode power handling, which is advantageous for achieving complete inter-modal conversion.

In conclusion, we propose a forward Brillouin scattering scheme for inter-modal conversion using quasi-phase-matching, achieved by periodically modulating the waveguide width along the propagation direction. This quasi-phase-matching approach enables effective acoustic mode confinement while satisfying the phase-matching condition. Through systematic optimization of the propagation direction, LN film thickness, etch depth, and waveguide width, we predict a complete conversion between optical modes can be realized by interacting with a \SI{1}{mW} acoustic mode over a waveguide length of \SI{1.1}{mm}. 

\smallskip{}

\begin{acknowledgments}
This work was funded by the National Natural Science Foundation of China (Grant Nos.~92265210,123B2068, 12374361 and 12293053), the Innovation Program for Quantum Science and Technology (Grant Nos.~2021ZD0300701 and 2024ZD0301500), the Strategic Priority Research Program of the Chinese Academy of Sciences (Grant No.~XDA0520503) and Industrial Prospect and Key Core Technology Projects of Jiangsu Provincial Key R \& D Program (BE2022071).. This work is also supported by the Fundamental Research Funds for the Central Universities and USTC Research Funds of the Double First-Class Initiative.
The numerical calculations are performed in the Supercomputing Center of USTC, and this work was partially carried out at the USTC Center for Micro and Nanoscale Research and Fabrication.
\end{acknowledgments}

\bibliographystyle{Zou}
\bibliography{ref}

\begin{thebibliography}{21}%
\makeatletter
\providecommand \@ifxundefined [1]{%
 \@ifx{#1\undefined}
}%
\providecommand \@ifnum [1]{%
 \ifnum #1\expandafter \@firstoftwo
 \else \expandafter \@secondoftwo
 \fi
}%
\providecommand \@ifx [1]{%
 \ifx #1\expandafter \@firstoftwo
 \else \expandafter \@secondoftwo
 \fi
}%
\providecommand \natexlab [1]{#1}%
\providecommand \enquote  [1]{``#1''}%
\providecommand \bibnamefont  [1]{#1}%
\providecommand \bibfnamefont [1]{#1}%
\providecommand \citenamefont [1]{#1}%
\providecommand \href@noop [0]{\@secondoftwo}%
\providecommand \href [0]{\begingroup \@sanitize@url \@href}%
\providecommand \@href[1]{\@@startlink{#1}\@@href}%
\providecommand \@@href[1]{\endgroup#1\@@endlink}%
\providecommand \@sanitize@url [0]{\catcode `\\12\catcode `\$12\catcode
  `\&12\catcode `\#12\catcode `\^12\catcode `\_12\catcode `\%12\relax}%
\providecommand \@@startlink[1]{}%
\providecommand \@@endlink[0]{}%
\providecommand \url  [0]{\begingroup\@sanitize@url \@url }%
\providecommand \@url [1]{\endgroup\@href {#1}{\urlprefix }}%
\providecommand \urlprefix  [0]{URL }%
\providecommand \Eprint [0]{\href }%
\providecommand \doibase [0]{http://dx.doi.org/}%
\providecommand \selectlanguage [0]{\@gobble}%
\providecommand \bibinfo  [0]{\@secondoftwo}%
\providecommand \bibfield  [0]{\@secondoftwo}%
\providecommand \translation [1]{[#1]}%
\providecommand \BibitemOpen [0]{}%
\providecommand \bibitemStop [0]{}%
\providecommand \bibitemNoStop [0]{.\EOS\space}%
\providecommand \EOS [0]{\spacefactor3000\relax}%
\providecommand \BibitemShut  [1]{\csname bibitem#1\endcsname}%
\let\auto@bib@innerbib\@empty
\bibitem [{\citenamefont {Wang}\ \emph {et~al.}(2018)\citenamefont {Wang},
  \citenamefont {Zhang}, \citenamefont {Chen}, \citenamefont {Bertrand},
  \citenamefont {{Shams-Ansari}}, \citenamefont {Chandrasekhar}, \citenamefont
  {Winzer},\ and\ \citenamefont {Lon{\v c}ar}}]{wang_Integrated_2018}%
  \BibitemOpen
  \bibfield  {author} {\bibinfo {author} {\bibfnamefont {C.}~\bibnamefont
  {Wang}}, \bibinfo {author} {\bibfnamefont {M.}~\bibnamefont {Zhang}},
  \bibinfo {author} {\bibfnamefont {X.}~\bibnamefont {Chen}}, \bibinfo {author}
  {\bibfnamefont {M.}~\bibnamefont {Bertrand}}, \bibinfo {author}
  {\bibfnamefont {A.}~\bibnamefont {{Shams-Ansari}}}, \bibinfo {author}
  {\bibfnamefont {S.}~\bibnamefont {Chandrasekhar}}, \bibinfo {author}
  {\bibfnamefont {P.}~\bibnamefont {Winzer}}, \ and\ \bibinfo {author}
  {\bibfnamefont {M.}~\bibnamefont {Lon{\v c}ar}},\ }\bibfield  {title}
  {\enquote {\bibinfo {title} {Integrated lithium niobate electro-optic
  modulators operating at {{CMOS-compatible}} voltages},}\ }\href {\doibase
  10.1038/s41586-018-0551-y} {\bibfield  {journal} {\bibinfo  {journal}
  {Nature}\ }\textbf {\bibinfo {volume} {562}},\ \bibinfo {pages} {101}
  (\bibinfo {year} {2018})}\BibitemShut {NoStop}%
\bibitem [{\citenamefont {Xue}\ \emph {et~al.}(2022)\citenamefont {Xue},
  \citenamefont {Gan}, \citenamefont {Chen}, \citenamefont {Chen},
  \citenamefont {Ruan}, \citenamefont {Zhang}, \citenamefont {Liu},
  \citenamefont {Dai}, \citenamefont {Guo},\ and\ \citenamefont
  {Liu}}]{xue_Breaking_2022}%
  \BibitemOpen
  \bibfield  {author} {\bibinfo {author} {\bibfnamefont {Y.}~\bibnamefont
  {Xue}}, \bibinfo {author} {\bibfnamefont {R.}~\bibnamefont {Gan}}, \bibinfo
  {author} {\bibfnamefont {K.}~\bibnamefont {Chen}}, \bibinfo {author}
  {\bibfnamefont {G.}~\bibnamefont {Chen}}, \bibinfo {author} {\bibfnamefont
  {Z.}~\bibnamefont {Ruan}}, \bibinfo {author} {\bibfnamefont {J.}~\bibnamefont
  {Zhang}}, \bibinfo {author} {\bibfnamefont {J.}~\bibnamefont {Liu}}, \bibinfo
  {author} {\bibfnamefont {D.}~\bibnamefont {Dai}}, \bibinfo {author}
  {\bibfnamefont {C.}~\bibnamefont {Guo}}, \ and\ \bibinfo {author}
  {\bibfnamefont {L.}~\bibnamefont {Liu}},\ }\bibfield  {title} {\enquote
  {\bibinfo {title} {Breaking the bandwidth limit of a high-quality-factor ring
  modulator based on thin-film lithium niobate},}\ }\href {\doibase
  10.1364/OPTICA.470596} {\bibfield  {journal} {\bibinfo  {journal} {Optica}\
  }\textbf {\bibinfo {volume} {9}},\ \bibinfo {pages} {1131} (\bibinfo {year}
  {2022})}\BibitemShut {NoStop}%
\bibitem [{\citenamefont {Sinatkas}\ \emph {et~al.}(2021)\citenamefont
  {Sinatkas}, \citenamefont {Christopoulos}, \citenamefont {Tsilipakos},\ and\
  \citenamefont {Kriezis}}]{sinatkas_Electrooptic_2021}%
  \BibitemOpen
  \bibfield  {author} {\bibinfo {author} {\bibfnamefont {G.}~\bibnamefont
  {Sinatkas}}, \bibinfo {author} {\bibfnamefont {T.}~\bibnamefont
  {Christopoulos}}, \bibinfo {author} {\bibfnamefont {O.}~\bibnamefont
  {Tsilipakos}}, \ and\ \bibinfo {author} {\bibfnamefont {E.~E.}\ \bibnamefont
  {Kriezis}},\ }\bibfield  {title} {\enquote {\bibinfo {title} {Electro-optic
  modulation in integrated photonics},}\ }\href {\doibase 10.1063/5.0048712}
  {\bibfield  {journal} {\bibinfo  {journal} {Journal of Applied Physics}\
  }\textbf {\bibinfo {volume} {130}},\ \bibinfo {pages} {010901} (\bibinfo
  {year} {2021})}\BibitemShut {NoStop}%
\bibitem [{\citenamefont {Watts}\ \emph {et~al.}(2013)\citenamefont {Watts},
  \citenamefont {Sun}, \citenamefont {DeRose}, \citenamefont {Trotter},
  \citenamefont {Young},\ and\ \citenamefont {Nielson}}]{watts_Adiabatic_2013}%
  \BibitemOpen
  \bibfield  {author} {\bibinfo {author} {\bibfnamefont {M.~R.}\ \bibnamefont
  {Watts}}, \bibinfo {author} {\bibfnamefont {J.}~\bibnamefont {Sun}}, \bibinfo
  {author} {\bibfnamefont {C.}~\bibnamefont {DeRose}}, \bibinfo {author}
  {\bibfnamefont {D.~C.}\ \bibnamefont {Trotter}}, \bibinfo {author}
  {\bibfnamefont {R.~W.}\ \bibnamefont {Young}}, \ and\ \bibinfo {author}
  {\bibfnamefont {G.~N.}\ \bibnamefont {Nielson}},\ }\bibfield  {title}
  {\enquote {\bibinfo {title} {Adiabatic thermo-optic {{Mach}}--{{Zehnder}}
  switch},}\ }\href {\doibase 10.1364/OL.38.000733} {\bibfield  {journal}
  {\bibinfo  {journal} {Optics Letters}\ }\textbf {\bibinfo {volume} {38}},\
  \bibinfo {pages} {733} (\bibinfo {year} {2013})}\BibitemShut {NoStop}%
\bibitem [{\citenamefont {Lu}\ \emph {et~al.}(2015)\citenamefont {Lu},
  \citenamefont {Murray}, \citenamefont {Jayatilleka},\ and\ \citenamefont
  {Chrostowski}}]{lu_Michelson_2015}%
  \BibitemOpen
  \bibfield  {author} {\bibinfo {author} {\bibfnamefont {Z.}~\bibnamefont
  {Lu}}, \bibinfo {author} {\bibfnamefont {K.}~\bibnamefont {Murray}}, \bibinfo
  {author} {\bibfnamefont {H.}~\bibnamefont {Jayatilleka}}, \ and\ \bibinfo
  {author} {\bibfnamefont {L.}~\bibnamefont {Chrostowski}},\ }\bibfield
  {title} {\enquote {\bibinfo {title} {Michelson {{Interferometer Thermo-Optic
  Switch}} on {{SOI With}} a 50-{{{\textmu}W Power Consumption}}},}\ }\href
  {\doibase 10.1109/LPT.2015.2462341} {\bibfield  {journal} {\bibinfo
  {journal} {IEEE Photonics Technology Letters}\ }\textbf {\bibinfo {volume}
  {27}},\ \bibinfo {pages} {2319} (\bibinfo {year} {2015})}\BibitemShut
  {NoStop}%
\bibitem [{\citenamefont {Yu}\ and\ \citenamefont
  {Qiu}(2022)}]{yu_Compact_2022}%
  \BibitemOpen
  \bibfield  {author} {\bibinfo {author} {\bibfnamefont {H.}~\bibnamefont
  {Yu}}\ and\ \bibinfo {author} {\bibfnamefont {F.}~\bibnamefont {Qiu}},\
  }\bibfield  {title} {\enquote {\bibinfo {title} {Compact thermo-optic
  modulator based on a titanium dioxide micro-ring resonator},}\ }\href
  {\doibase 10.1364/OL.456876} {\bibfield  {journal} {\bibinfo  {journal}
  {Optics Letters}\ }\textbf {\bibinfo {volume} {47}},\ \bibinfo {pages} {2093}
  (\bibinfo {year} {2022})}\BibitemShut {NoStop}%
\bibitem [{\citenamefont {Zhang}\ \emph
  {et~al.}(2025{\natexlab{a}})\citenamefont {Zhang}, \citenamefont {Xue},
  \citenamefont {Chen}, \citenamefont {Guo}, \citenamefont {Wang},
  \citenamefont {Li},\ and\ \citenamefont {Yan}}]{zhang_Highly_2025}%
  \BibitemOpen
  \bibfield  {author} {\bibinfo {author} {\bibfnamefont {L.}~\bibnamefont
  {Zhang}}, \bibinfo {author} {\bibfnamefont {Y.}~\bibnamefont {Xue}}, \bibinfo
  {author} {\bibfnamefont {Z.}~\bibnamefont {Chen}}, \bibinfo {author}
  {\bibfnamefont {Y.}~\bibnamefont {Guo}}, \bibinfo {author} {\bibfnamefont
  {J.}~\bibnamefont {Wang}}, \bibinfo {author} {\bibfnamefont {J.}~\bibnamefont
  {Li}}, \ and\ \bibinfo {author} {\bibfnamefont {J.}~\bibnamefont {Yan}},\
  }\bibfield  {title} {\enquote {\bibinfo {title} {Highly {{Efficient
  Acousto-Optic Modulation Driven}} by {{Ultra-Low Power}} in {{Integrated
  Photonic}}--{{Phononic Waveguides}}},}\ }\href {\doibase
  10.1002/lpor.202401952} {\bibfield  {journal} {\bibinfo  {journal} {Laser \&
  Photonics Reviews}\ }\textbf {\bibinfo {volume} {n/a}},\ \bibinfo {pages}
  {2401952} (\bibinfo {year} {2025}{\natexlab{a}})}\BibitemShut {NoStop}%
\bibitem [{\citenamefont {Xu}\ \emph {et~al.}(2022)\citenamefont {Xu},
  \citenamefont {Wang}, \citenamefont {Yang}, \citenamefont {Wang},
  \citenamefont {Zhang}, \citenamefont {Wang}, \citenamefont {Dong},
  \citenamefont {Sun}, \citenamefont {Guo},\ and\ \citenamefont
  {Zou}}]{xu_Highfrequency_2022}%
  \BibitemOpen
  \bibfield  {author} {\bibinfo {author} {\bibfnamefont {X.-B.}\ \bibnamefont
  {Xu}}, \bibinfo {author} {\bibfnamefont {J.-Q.}\ \bibnamefont {Wang}},
  \bibinfo {author} {\bibfnamefont {Y.-H.}\ \bibnamefont {Yang}}, \bibinfo
  {author} {\bibfnamefont {W.}~\bibnamefont {Wang}}, \bibinfo {author}
  {\bibfnamefont {Y.-L.}\ \bibnamefont {Zhang}}, \bibinfo {author}
  {\bibfnamefont {B.-Z.}\ \bibnamefont {Wang}}, \bibinfo {author}
  {\bibfnamefont {C.-H.}\ \bibnamefont {Dong}}, \bibinfo {author}
  {\bibfnamefont {L.}~\bibnamefont {Sun}}, \bibinfo {author} {\bibfnamefont
  {G.-C.}\ \bibnamefont {Guo}}, \ and\ \bibinfo {author} {\bibfnamefont
  {C.-L.}\ \bibnamefont {Zou}},\ }\bibfield  {title} {\enquote {\bibinfo
  {title} {High-frequency traveling-wave phononic cavity with sub-micron
  wavelength},}\ }\href {\doibase 10.1063/5.0086751} {\bibfield  {journal}
  {\bibinfo  {journal} {Applied Physics Letters}\ }\textbf {\bibinfo {volume}
  {120}},\ \bibinfo {pages} {163503} (\bibinfo {year} {2022})}\BibitemShut
  {NoStop}%
\bibitem [{\citenamefont {Liu}\ \emph {et~al.}(2019)\citenamefont {Liu},
  \citenamefont {Li},\ and\ \citenamefont {Li}}]{liu_Electromechanical_2019}%
  \BibitemOpen
  \bibfield  {author} {\bibinfo {author} {\bibfnamefont {Q.}~\bibnamefont
  {Liu}}, \bibinfo {author} {\bibfnamefont {H.}~\bibnamefont {Li}}, \ and\
  \bibinfo {author} {\bibfnamefont {M.}~\bibnamefont {Li}},\ }\bibfield
  {title} {\enquote {\bibinfo {title} {Electromechanical {{Brillouin}}
  scattering in integrated optomechanical waveguides},}\ }\href {\doibase
  10.1364/OPTICA.6.000778} {\bibfield  {journal} {\bibinfo  {journal} {Optica}\
  }\textbf {\bibinfo {volume} {6}},\ \bibinfo {pages} {778} (\bibinfo {year}
  {2019})}\BibitemShut {NoStop}%
\bibitem [{\citenamefont {Kittlaus}\ \emph {et~al.}(2021)\citenamefont
  {Kittlaus}, \citenamefont {Jones}, \citenamefont {Rakich}, \citenamefont
  {Otterstrom}, \citenamefont {Muller},\ and\ \citenamefont
  {{Rais-Zadeh}}}]{kittlaus_Electrically_2021}%
  \BibitemOpen
  \bibfield  {author} {\bibinfo {author} {\bibfnamefont {E.~A.}\ \bibnamefont
  {Kittlaus}}, \bibinfo {author} {\bibfnamefont {W.~M.}\ \bibnamefont {Jones}},
  \bibinfo {author} {\bibfnamefont {P.~T.}\ \bibnamefont {Rakich}}, \bibinfo
  {author} {\bibfnamefont {N.~T.}\ \bibnamefont {Otterstrom}}, \bibinfo
  {author} {\bibfnamefont {R.~E.}\ \bibnamefont {Muller}}, \ and\ \bibinfo
  {author} {\bibfnamefont {M.}~\bibnamefont {{Rais-Zadeh}}},\ }\bibfield
  {title} {\enquote {\bibinfo {title} {Electrically driven acousto-optics and
  broadband non-reciprocity in silicon photonics},}\ }\href {\doibase
  10.1038/s41566-020-00711-9} {\bibfield  {journal} {\bibinfo  {journal}
  {Nature Photonics}\ }\textbf {\bibinfo {volume} {15}},\ \bibinfo {pages} {43}
  (\bibinfo {year} {2021})}\BibitemShut {NoStop}%
\bibitem [{\citenamefont {Sarabalis}\ \emph {et~al.}(2020)\citenamefont
  {Sarabalis}, \citenamefont {McKenna}, \citenamefont {Patel}, \citenamefont
  {Van~Laer},\ and\ \citenamefont
  {{Safavi-Naeini}}}]{sarabalis_Acoustooptic_2020}%
  \BibitemOpen
  \bibfield  {author} {\bibinfo {author} {\bibfnamefont {C.~J.}\ \bibnamefont
  {Sarabalis}}, \bibinfo {author} {\bibfnamefont {T.~P.}\ \bibnamefont
  {McKenna}}, \bibinfo {author} {\bibfnamefont {R.~N.}\ \bibnamefont {Patel}},
  \bibinfo {author} {\bibfnamefont {R.}~\bibnamefont {Van~Laer}}, \ and\
  \bibinfo {author} {\bibfnamefont {A.~H.}\ \bibnamefont {{Safavi-Naeini}}},\
  }\bibfield  {title} {\enquote {\bibinfo {title} {Acousto-optic modulation in
  lithium niobate on sapphire},}\ }\href {\doibase 10.1063/5.0012288}
  {\bibfield  {journal} {\bibinfo  {journal} {APL Photonics}\ }\textbf
  {\bibinfo {volume} {5}},\ \bibinfo {pages} {086104} (\bibinfo {year}
  {2020})}\BibitemShut {NoStop}%
\bibitem [{\citenamefont {Zhang}\ \emph
  {et~al.}(2025{\natexlab{b}})\citenamefont {Zhang}, \citenamefont {Zeng},
  \citenamefont {Qin}, \citenamefont {Yang}, \citenamefont {Tian},
  \citenamefont {Wang}, \citenamefont {Dong}, \citenamefont {Xu}, \citenamefont
  {Ye}, \citenamefont {Guo},\ and\ \citenamefont {Zou}}]{zhang_Onchip_2025}%
  \BibitemOpen
  \bibfield  {author} {\bibinfo {author} {\bibfnamefont {J.-Z.}\ \bibnamefont
  {Zhang}}, \bibinfo {author} {\bibfnamefont {Y.}~\bibnamefont {Zeng}},
  \bibinfo {author} {\bibfnamefont {Q.}~\bibnamefont {Qin}}, \bibinfo {author}
  {\bibfnamefont {Y.-H.}\ \bibnamefont {Yang}}, \bibinfo {author}
  {\bibfnamefont {Z.-H.}\ \bibnamefont {Tian}}, \bibinfo {author}
  {\bibfnamefont {J.-Q.}\ \bibnamefont {Wang}}, \bibinfo {author}
  {\bibfnamefont {C.-H.}\ \bibnamefont {Dong}}, \bibinfo {author}
  {\bibfnamefont {X.-B.}\ \bibnamefont {Xu}}, \bibinfo {author} {\bibfnamefont
  {M.-Y.}\ \bibnamefont {Ye}}, \bibinfo {author} {\bibfnamefont {G.-C.}\
  \bibnamefont {Guo}}, \ and\ \bibinfo {author} {\bibfnamefont {C.-L.}\
  \bibnamefont {Zou}},\ }\bibfield  {title} {\enquote {\bibinfo {title}
  {On-chip 7 {{GHz}} acousto-optic modulators for visible wavelengths},}\
  }\href {\doibase 10.1364/OE.540356} {\bibfield  {journal} {\bibinfo
  {journal} {Optics Express}\ }\textbf {\bibinfo {volume} {33}},\ \bibinfo
  {pages} {5562} (\bibinfo {year} {2025}{\natexlab{b}})}\BibitemShut {NoStop}%
\bibitem [{\citenamefont {Sarabalis}\ \emph {et~al.}(2021)\citenamefont
  {Sarabalis}, \citenamefont {Laer}, \citenamefont {Patel}, \citenamefont
  {Dahmani}, \citenamefont {Jiang}, \citenamefont {Mayor},\ and\ \citenamefont
  {{Safavi-Naeini}}}]{sarabalis_Acoustooptic_2021}%
  \BibitemOpen
  \bibfield  {author} {\bibinfo {author} {\bibfnamefont {C.~J.}\ \bibnamefont
  {Sarabalis}}, \bibinfo {author} {\bibfnamefont {R.~V.}\ \bibnamefont {Laer}},
  \bibinfo {author} {\bibfnamefont {R.~N.}\ \bibnamefont {Patel}}, \bibinfo
  {author} {\bibfnamefont {Y.~D.}\ \bibnamefont {Dahmani}}, \bibinfo {author}
  {\bibfnamefont {W.}~\bibnamefont {Jiang}}, \bibinfo {author} {\bibfnamefont
  {F.~M.}\ \bibnamefont {Mayor}}, \ and\ \bibinfo {author} {\bibfnamefont
  {A.~H.}\ \bibnamefont {{Safavi-Naeini}}},\ }\bibfield  {title} {\enquote
  {\bibinfo {title} {Acousto-optic modulation of a wavelength-scale
  waveguide},}\ }\href {\doibase 10.1364/OPTICA.413401} {\bibfield  {journal}
  {\bibinfo  {journal} {Optica}\ }\textbf {\bibinfo {volume} {8}},\ \bibinfo
  {pages} {477} (\bibinfo {year} {2021})}\BibitemShut {NoStop}%
\bibitem [{\citenamefont {Xu}\ \emph {et~al.}(2025)\citenamefont {Xu},
  \citenamefont {Zhu}, \citenamefont {Yang}, \citenamefont {Wang},
  \citenamefont {Zeng}, \citenamefont {Zou}, \citenamefont {Lu}, \citenamefont
  {Zhang}, \citenamefont {Wang}, \citenamefont {Guo}, \citenamefont {Sun},\
  and\ \citenamefont {Zou}}]{Xu2025}%
  \BibitemOpen
  \bibfield  {author} {\bibinfo {author} {\bibfnamefont {X.-B.}\ \bibnamefont
  {Xu}}, \bibinfo {author} {\bibfnamefont {Z.-X.}\ \bibnamefont {Zhu}},
  \bibinfo {author} {\bibfnamefont {Y.-H.}\ \bibnamefont {Yang}}, \bibinfo
  {author} {\bibfnamefont {J.-Q.}\ \bibnamefont {Wang}}, \bibinfo {author}
  {\bibfnamefont {Y.}~\bibnamefont {Zeng}}, \bibinfo {author} {\bibfnamefont
  {J.-H.}\ \bibnamefont {Zou}}, \bibinfo {author} {\bibfnamefont
  {J.}~\bibnamefont {Lu}}, \bibinfo {author} {\bibfnamefont {Y.-L.}\
  \bibnamefont {Zhang}}, \bibinfo {author} {\bibfnamefont {W.}~\bibnamefont
  {Wang}}, \bibinfo {author} {\bibfnamefont {G.-C.}\ \bibnamefont {Guo}},
  \bibinfo {author} {\bibfnamefont {L.}~\bibnamefont {Sun}}, \ and\ \bibinfo
  {author} {\bibfnamefont {C.-L.}\ \bibnamefont {Zou}},\ }\bibfield  {title}
  {\enquote {\bibinfo {title} {{Magnetic-free optical mode degeneracy lifting
  in lithium niobate microring resonators}},}\ }\href
  {http://arxiv.org/abs/2509.01940} {\bibfield  {journal} {\bibinfo  {journal}
  {arXiv preprint}\ ,\ \bibinfo {pages} {2509.01940}} (\bibinfo {year}
  {2025})}\BibitemShut {NoStop}%
\bibitem [{\citenamefont {Yang}\ \emph {et~al.}(2025)\citenamefont {Yang},
  \citenamefont {Wang}, \citenamefont {Zhu}, \citenamefont {Zeng},
  \citenamefont {Li}, \citenamefont {Zhang}, \citenamefont {Lu}, \citenamefont
  {Zhang}, \citenamefont {Wang}, \citenamefont {Dong}, \citenamefont {Xu},
  \citenamefont {Guo}, \citenamefont {Sun},\ and\ \citenamefont
  {Zou}}]{Yang2025}%
  \BibitemOpen
  \bibfield  {author} {\bibinfo {author} {\bibfnamefont {Y.-H.}\ \bibnamefont
  {Yang}}, \bibinfo {author} {\bibfnamefont {J.-Q.}\ \bibnamefont {Wang}},
  \bibinfo {author} {\bibfnamefont {Z.-X.}\ \bibnamefont {Zhu}}, \bibinfo
  {author} {\bibfnamefont {Y.}~\bibnamefont {Zeng}}, \bibinfo {author}
  {\bibfnamefont {M.}~\bibnamefont {Li}}, \bibinfo {author} {\bibfnamefont
  {Y.-L.}\ \bibnamefont {Zhang}}, \bibinfo {author} {\bibfnamefont
  {J.}~\bibnamefont {Lu}}, \bibinfo {author} {\bibfnamefont {Q.}~\bibnamefont
  {Zhang}}, \bibinfo {author} {\bibfnamefont {W.}~\bibnamefont {Wang}},
  \bibinfo {author} {\bibfnamefont {C.-H.}\ \bibnamefont {Dong}}, \bibinfo
  {author} {\bibfnamefont {X.-B.}\ \bibnamefont {Xu}}, \bibinfo {author}
  {\bibfnamefont {G.-C.}\ \bibnamefont {Guo}}, \bibinfo {author} {\bibfnamefont
  {L.}~\bibnamefont {Sun}}, \ and\ \bibinfo {author} {\bibfnamefont {C.-L.}\
  \bibnamefont {Zou}},\ }\bibfield  {title} {\enquote {\bibinfo {title}
  {{Multi-Channel Microwave-to-Optics Conversion Utilizing a Hybrid
  Photonic-Phononic Waveguide}},}\ }\href {http://arxiv.org/abs/2509.10052}
  {\bibfield  {journal} {\bibinfo  {journal} {arXiv preprint}\ ,\ \bibinfo
  {pages} {2509.10052}} (\bibinfo {year} {2025})}\BibitemShut {NoStop}%
\bibitem [{\citenamefont {Lu}\ \emph {et~al.}(2024)\citenamefont {Lu},
  \citenamefont {Chang}, \citenamefont {Tran}, \citenamefont {Komljenovic},
  \citenamefont {Bowers},\ and\ \citenamefont {Srinivasan}}]{Lu2024}%
  \BibitemOpen
  \bibfield  {author} {\bibinfo {author} {\bibfnamefont {X.}~\bibnamefont
  {Lu}}, \bibinfo {author} {\bibfnamefont {L.}~\bibnamefont {Chang}}, \bibinfo
  {author} {\bibfnamefont {M.~A.}\ \bibnamefont {Tran}}, \bibinfo {author}
  {\bibfnamefont {T.}~\bibnamefont {Komljenovic}}, \bibinfo {author}
  {\bibfnamefont {J.~E.}\ \bibnamefont {Bowers}}, \ and\ \bibinfo {author}
  {\bibfnamefont {K.}~\bibnamefont {Srinivasan}},\ }\bibfield  {title}
  {\enquote {\bibinfo {title} {{Emerging integrated laser technologies in the
  visible and short near-infrared regimes}},}\ }\href {\doibase
  10.1038/s41566-024-01529-5} {\bibfield  {journal} {\bibinfo  {journal}
  {Nature Photonics}\ }\textbf {\bibinfo {volume} {18}},\ \bibinfo {pages}
  {1010} (\bibinfo {year} {2024})}\BibitemShut {NoStop}%
\bibitem [{\citenamefont {Browaeys}\ and\ \citenamefont
  {Lahaye}(2020)}]{Browaeys2020}%
  \BibitemOpen
  \bibfield  {author} {\bibinfo {author} {\bibfnamefont {A.}~\bibnamefont
  {Browaeys}}\ and\ \bibinfo {author} {\bibfnamefont {T.}~\bibnamefont
  {Lahaye}},\ }\bibfield  {title} {\enquote {\bibinfo {title} {{Many-body
  physics with individually controlled Rydberg atoms}},}\ }\href {\doibase
  10.1038/s41567-019-0733-z} {\bibfield  {journal} {\bibinfo  {journal} {Nature
  Physics}\ }\textbf {\bibinfo {volume} {16}},\ \bibinfo {pages} {132}
  (\bibinfo {year} {2020})}\BibitemShut {NoStop}%
\bibitem [{\citenamefont {Yang}\ \emph {et~al.}(2024)\citenamefont {Yang},
  \citenamefont {Wang}, \citenamefont {Xu}, \citenamefont {Li}, \citenamefont
  {Zhang}, \citenamefont {Pan}, \citenamefont {Xiao}, \citenamefont {Wang},
  \citenamefont {Guo}, \citenamefont {Sun},\ and\ \citenamefont
  {Zou}}]{yang_Proposal_2024}%
  \BibitemOpen
  \bibfield  {author} {\bibinfo {author} {\bibfnamefont {Y.-H.}\ \bibnamefont
  {Yang}}, \bibinfo {author} {\bibfnamefont {J.-Q.}\ \bibnamefont {Wang}},
  \bibinfo {author} {\bibfnamefont {X.-B.}\ \bibnamefont {Xu}}, \bibinfo
  {author} {\bibfnamefont {M.}~\bibnamefont {Li}}, \bibinfo {author}
  {\bibfnamefont {Y.-L.}\ \bibnamefont {Zhang}}, \bibinfo {author}
  {\bibfnamefont {X.}~\bibnamefont {Pan}}, \bibinfo {author} {\bibfnamefont
  {L.}~\bibnamefont {Xiao}}, \bibinfo {author} {\bibfnamefont {W.}~\bibnamefont
  {Wang}}, \bibinfo {author} {\bibfnamefont {G.-c.}\ \bibnamefont {Guo}},
  \bibinfo {author} {\bibfnamefont {L.}~\bibnamefont {Sun}}, \ and\ \bibinfo
  {author} {\bibfnamefont {C.-l.}\ \bibnamefont {Zou}},\ }\bibfield  {title}
  {\enquote {\bibinfo {title} {Proposal for {{Brillouin}} microwave-to-optical
  conversion on a chip [{{Invited}}]},}\ }\href {\doibase 10.1364/OME.534817}
  {\bibfield  {journal} {\bibinfo  {journal} {Optical Materials Express}\
  }\textbf {\bibinfo {volume} {14}},\ \bibinfo {pages} {2400} (\bibinfo {year}
  {2024})}\BibitemShut {NoStop}%
\bibitem [{\citenamefont {Zhang}\ \emph {et~al.}(2024)\citenamefont {Zhang},
  \citenamefont {Cui}, \citenamefont {Chen},\ and\ \citenamefont
  {Fan}}]{zhang_Integratedwaveguidebased_2024}%
  \BibitemOpen
  \bibfield  {author} {\bibinfo {author} {\bibfnamefont {L.}~\bibnamefont
  {Zhang}}, \bibinfo {author} {\bibfnamefont {C.}~\bibnamefont {Cui}}, \bibinfo
  {author} {\bibfnamefont {P.-K.}\ \bibnamefont {Chen}}, \ and\ \bibinfo
  {author} {\bibfnamefont {L.}~\bibnamefont {Fan}},\ }\bibfield  {title}
  {\enquote {\bibinfo {title} {Integrated-waveguide-based acousto-optic
  modulation with complete optical conversion},}\ }\href {\doibase
  10.1364/OPTICA.488271} {\bibfield  {journal} {\bibinfo  {journal} {Optica}\
  }\textbf {\bibinfo {volume} {11}},\ \bibinfo {pages} {184} (\bibinfo {year}
  {2024})}\BibitemShut {NoStop}%
\bibitem [{\citenamefont {Yang}\ \emph {et~al.}(2023)\citenamefont {Yang},
  \citenamefont {Wang}, \citenamefont {Zhu}, \citenamefont {Xu}, \citenamefont
  {Zhang}, \citenamefont {Lu}, \citenamefont {Zeng}, \citenamefont {Dong},
  \citenamefont {Sun}, \citenamefont {Guo},\ and\ \citenamefont
  {Zou}}]{yang_Stimulated_2023}%
  \BibitemOpen
  \bibfield  {author} {\bibinfo {author} {\bibfnamefont {Y.-H.}\ \bibnamefont
  {Yang}}, \bibinfo {author} {\bibfnamefont {J.-Q.}\ \bibnamefont {Wang}},
  \bibinfo {author} {\bibfnamefont {Z.-X.}\ \bibnamefont {Zhu}}, \bibinfo
  {author} {\bibfnamefont {X.-B.}\ \bibnamefont {Xu}}, \bibinfo {author}
  {\bibfnamefont {Q.}~\bibnamefont {Zhang}}, \bibinfo {author} {\bibfnamefont
  {J.}~\bibnamefont {Lu}}, \bibinfo {author} {\bibfnamefont {Y.}~\bibnamefont
  {Zeng}}, \bibinfo {author} {\bibfnamefont {C.-H.}\ \bibnamefont {Dong}},
  \bibinfo {author} {\bibfnamefont {L.}~\bibnamefont {Sun}}, \bibinfo {author}
  {\bibfnamefont {G.-C.}\ \bibnamefont {Guo}}, \ and\ \bibinfo {author}
  {\bibfnamefont {C.-L.}\ \bibnamefont {Zou}},\ }\bibfield  {title} {\enquote
  {\bibinfo {title} {Stimulated {{Brillouin}} interaction between guided
  phonons and photons in a lithium niobate waveguide},}\ }\href {\doibase
  10.1007/s11433-023-2272-y} {\bibfield  {journal} {\bibinfo  {journal}
  {Science China Physics, Mechanics \& Astronomy}\ }\textbf {\bibinfo {volume}
  {67}},\ \bibinfo {pages} {214221} (\bibinfo {year} {2023})}\BibitemShut
  {NoStop}%
\bibitem [{\citenamefont {Ogi}\ \emph {et~al.}(2002)\citenamefont {Ogi},
  \citenamefont {Kawasaki}, \citenamefont {Hirao},\ and\ \citenamefont
  {Ledbetter}}]{ogi_Acoustic_2002}%
  \BibitemOpen
  \bibfield  {author} {\bibinfo {author} {\bibfnamefont {H.}~\bibnamefont
  {Ogi}}, \bibinfo {author} {\bibfnamefont {Y.}~\bibnamefont {Kawasaki}},
  \bibinfo {author} {\bibfnamefont {M.}~\bibnamefont {Hirao}}, \ and\ \bibinfo
  {author} {\bibfnamefont {H.}~\bibnamefont {Ledbetter}},\ }\bibfield  {title}
  {\enquote {\bibinfo {title} {Acoustic spectroscopy of lithium niobate:
  {{Elastic}} and piezoelectric coefficients},}\ }\href {\doibase
  10.1063/1.1497702} {\bibfield  {journal} {\bibinfo  {journal} {Journal of
  Applied Physics}\ }\textbf {\bibinfo {volume} {92}},\ \bibinfo {pages} {2451}
  (\bibinfo {year} {2002})}\BibitemShut {NoStop}%
\end{thebibliography}%

\end{document}